\documentclass[%
 reprint,
 amsmath,amssymb,
 aps,
prb,
 longbibliography
]{revtex4-2}

\usepackage{graphicx}
\usepackage{dcolumn}
\usepackage{bm}
\usepackage{physics}
\usepackage{mleftright}
\usepackage{siunitx}
\usepackage{makecell}
\usepackage{hyperref}

\usepackage{ulem}
\usepackage[usenames]{xcolor}
\usepackage{epstopdf}

\definecolor{mygreen}{RGB}{0,204,102}

\DeclareMathAlphabet{\mathpzc}{OT1}{pzc}{m}{it}

\begin{document}

\preprint{APS/123-QED}

\title{Vortex creation and control in the Kitaev spin liquid by local bond modulations}

\author{Seong-Hoon Jang}
\author{Yasuyuki Kato}
\author{Yukitoshi Motome}
\affiliation{
 Department of Applied Physics, The University of Tokyo, Tokyo 113-8656, Japan
}

\date{\today}

\begin{abstract}
The Kitaev model realizes a quantum spin liquid where the spin excitations are fractionalized into itinerant Majorana fermions and localized $\mathbb{Z}_2$ vortices. Quantum entanglement between the fractional excitations can be utilized for decoherence-free topological quantum computation. Of particular interest is the anyonic statistics realized by braiding the vortex excitations under a magnetic field. Despite the promising potential, the practical methodology for creation and control of the vortex excitations remains elusive thus far. Here we theoretically propose how one can create and move the vortices in the Kitaev spin liquid. We find that the vortices are induced by a local modulation of the exchange interaction; especially, the local Dzyaloshinskii-Moriya (symmetric off-diagonal) interaction can create vortices most efficiently in the (anti)ferromagnetic Kitaev model, as it effectively flips the sign of the Kitaev interaction. We test this idea by performing the {\it ab initio} calculation for a candidate material $\alpha$-RuCl$_3$ through the manipulation of the ligand positions that breaks the inversion symmetry and induces the local Dzyaloshinskii-Moriya interaction. We also demonstrate a braiding of vortices by adiabatically and successively changing the local bond modulations. 
\end{abstract}

\pacs{Valid PACS appear here}
\maketitle

\section{Introduction} 
\label{sec:introduction}

\begin{figure}
\includegraphics[width=1.0\columnwidth]{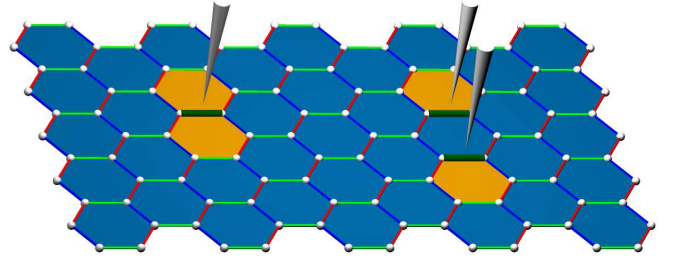} 
\caption{\label{fig:vortexcreation}
Schematic of the Kitaev model and vortices. The blue, red, and green lines denote the $x$, $y$, and $z$ bonds in Eq.~(\ref{eq:H_Kitaev}), respectively. 
The yellow hexagons represent the vortices induced by modulations on the thick green bonds from the vortex-free background denoted by the blue hexagons. The gray needles schematically represent local probes to induce the bond modulations.}
\end{figure}

The quantum spin liquid (QSL) is an exotic phase in which localized spins remain magnetically disordered, releasing their entropy down to zero temperature without breaking any symmetry of the system. Since the theoretical proposal by P. W. Anderson~\cite{AN1973}, a considerable amount of theoretical and experimental studies have been devoted to the search for the QSL~\cite{BA2010, ZH2017A, SA2017}. One of the recent breakthroughs in this situation has been made by a finding of an exactly solvable spin model by A. Kitaev~\cite{KI2006B}. The model adopts localized spin-$1/2$ moments on a honeycomb lattice with bond-dependent Ising-type interactions, whose competition prevents the long-range magnetic order down to zero temperature. The Hamiltonian is given as
\begin{equation}
\mathpzc{H}= \sum_{\mu}\sum_{\langle i,i' \rangle_\mu}K^\mu S_i^\mu S_{i'}^\mu,
\label{eq:H_Kitaev}
\end{equation} 
where $K^\mu$ denotes the coupling constant between the neighboring spins on the $\mu$($=x,y,z$) bonds, and $S_i^\mu$ represents the $\mu$ component of spin-$1/2$ operator at site $i$; see Fig.~\ref{fig:vortexcreation}. It was shown that the Kitaev model is exactly solvable by replacing the spin operators by Majorana fermion operators, and the ground state is shown to be a QSL. After it was pointed out that the Kitaev-type interactions can be achieved in a class of spin-orbit coupled Mott insulators~\cite{JA2009}, this exact solution has stimulated a fierce race for the materialization of the Kitaev QSL~\cite{TR2017, WI2017, HE2018, KN2019, TA2019, MO2020}. 

The QSL has a striking feature in its quasiparticle excitations; the quasiparticles are emergent from fractionalization of the quantum spins, and hence, they are intrinsically nonlocal and quantum entangled~\cite{BA2010, SA2017, ZH2017A}. Since the emergent quasiparticles created and annihilated in a pairwise fashion obey abelian or non-abelian statistics, their algebraic properties can be utilized for constructing fusion rules, braiding rules, and eventually quantum logic gates~\cite{SA1992, KI2003, NA2008, KI2006B}. As the massive entanglement under the topological order is robust against local perturbations, the interesting features of the QSL pave the way to decoherence-free topological quantum computation. The Kitaev model provides an excellent  platform for this possibility. The exact solution of this model indicates that the spin excitations are fractionalized into two types of quasiparticles, itinerant Majorana fermions and localized $\mathbb{Z}_2$ vortices. The latter quasiparticle is defined by the shortest Wilson loop on each hexagonal plaquette as 
\begin{equation}
\mathcal{W}_p= 2^6 \prod_{i\in p} S_{i}^{\bar{\mu}_{i}}, 
\label{eq:Wp}
\end{equation} 
where the product is taken for the six sites along the plaquette $p$; $\bar{\mu}_i$ denotes the type of the bond connected to the site $i$ from the outside of $p$. The ground-state QSL for the Kitaev model is given as the vortex-free state where all $\langle \mathcal{W}_p \rangle$ take $+1$. In the analogous of the Moore-Read state~\cite{MOORE1991362, 10.1143/PTPS.107.157}, Kitaev carefully examined that the decoherence-free topological quantum computation is made possible with the non-abelian statistics by introducing vortices $\langle \mathcal{W}_p \rangle=-1$ in the presence of a weak magnetic field~\cite{KI2006B}. Notably, the recent experimental discovery of the half-quantized thermal Hall conductivity in a candidate material $\alpha$-RuCl$_3$ supports the existence of such fractional quasiparticles~\cite{KA2018B,Yokoi2020preprint}, which allows us to expect the realization of the topological quantum computation in the Kitaev QSL. 

Nevertheless, an attempt to create and control the vortices in the Kitaev QSL has not been explored extensively yet in both experiment and theory. Still, some clues were given in the recent advances, e.g., the observation of the Abrikosov vortices in superconductors~\cite{PhysRevLett.62.214, PhysRevLett.64.2711, PhysRevLett.75.2754, PhysRevLett.78.4273, Troyanovski1999, Hoffman466, PhysRevLett.89.187003, PhysRevB.68.140503, PhysRevLett.96.097006, Guillamon2009}, the Majorana zero modes residing in topological superconductors~\cite{PhysRevLett.114.017001, Sun2017, Machida2019, Liu2019} and one-dimensional nanowires~\cite{Kitaev_2001, Alicea_2012, Leijnse_2012, Mourik2012, Beenakker2013, Lutchyn2018, Rahmani_2019}, and anyonic braiding statistics in two-dimensional electron systems confined in a heterostructure~\cite{Nakamura2020}. Despite recent proposals for controlling or detecting the fractional excitations in the Kitaev QSL by local probes like the scanning tunneling microscopy and the atomic force microscopy~\cite{PhysRevB.102.085412, PhysRevB.102.134423, PhysRevLett.125.227202, PhysRevLett.125.267206, PhysRevLett.126.127201}, further analyses are highly desired for the practical realization of fusion and braiding vortices excited under nonlocal quantum entanglement. 

In this paper, we propose an efficient way to create and control the vortices in the Kitaev QSL by using local modulations of the exchange interactions. First, we study the effect of various local bond modulations on the Kitaev model, and show that the local Dzyaloshinskii-Moriya (DM) interaction can create vortices most efficiently as it effectively flips the sign of the Kitaev interaction coefficient. Then, we test this idea by using the {\it ab initio} calculation for the candidate material $\alpha$-RuCl$_3$, and show that a manipulation of a ligand position that breaks the inversion symmetry induces the local DM interaction and induces a vortex-like feature in the field-induced QSL state. In addition, we demonstrate an adiabatic braiding process of vortices by successively applying the local bond modulations. 

The organization of the rest of this paper is as follows. In Sec.~\ref{sec:lvortex}, we examine the effect of local bond modulations on the Kitaev QSL. In Sec.~\ref{subsec:lbond}, we investigate the ground state of the Kitaev model when the Kitaev interaction on one of the bonds in the entire system is replaced by other interactions, by using the exact diagonalization. We show that the replacement by the DM interaction flips the sign of the vortices around the modulated bond most efficiently. In Sec.~\ref{subsec:caseRuCl3}, to illustrate the idea in a realistic situation, we perform the {\it ab initio} study for the effect of the bond modulation by ligand displacement in the Kitaev candidate material $\alpha$-RuCl$_3$. By estimating the modulations of the effective exchange interactions and performing the exact diagonalization of the modulated spin Hamiltonian, we show that a vortex-like feature is induced by the ligand displacement in the QSL phase in a magnetic field. In Sec.~\ref{sec:ccvortex}, we propose adiabatic processes for manipulating the positions of vortices, which make a braiding of a pair of vortices feasible. Section~\ref{sec:summary} is devoted to the summary. 

\section{Local vortex excitation} 
\label{sec:lvortex}

\subsection{Local bond modulation} 
\label{subsec:lbond}

We examine the vortex configuration in the ground state by modulating the exchange interaction in the Kitaev model in Eq.~(\ref{eq:H_Kitaev}). Specifically, we introduce a local bond modulation by replacing the Hamiltonian on a $z$ bond, $K^z S_i^z S_{i'}^z$, with 
\begin{equation}
\mathpzc{H}^{ \left( z \right) }_{ii'}=
\mathbf{S}_i^{\rm T}
\begin{bmatrix} 
J & \Gamma + D & \Gamma' \\
\Gamma - D & J & \Gamma' \\
\Gamma' & \Gamma' & J+K^z
\end{bmatrix}
\mathbf{S}_{i'},
\label{eq:Heff}
\end{equation}
where $J$, $D$, $\Gamma$, and $\Gamma'$ denote the isotropic Heisenberg, the asymmetric DM, and two types of the symmetric off-diagonal interactions, respectively; $\mathbf{S}_i=(S_i^x,S_i^y,S_i^z)^{\textrm{T}}$. In this subsection, on the target bond, we set $K^z=0$ and examine the effect of $J$, $D$, $\Gamma$, and $\Gamma'$ one by one; we take the ferromagnetic (FM) Kitaev interaction $K^x=K^y=K^z=K=-1$ on the other bonds [we will comment on the antiferromagnetic (AFM) case in the end of this subsection]. We use the exact diagonalization by the locally optimal block conjugate gradient (LOBCG) method~\cite{Knyazev2001} for finite-size clusters under the periodic boundary condition. To resolve the cumbersome degeneracy inherited to small-size clusters, we take $K$ randomly in the range of $[-1 - \varepsilon, -1 + \varepsilon]$ with $\varepsilon = 10^{-9}$, except for the target $z$ bond. 

\begin{figure}
\includegraphics[width=1.0\columnwidth]{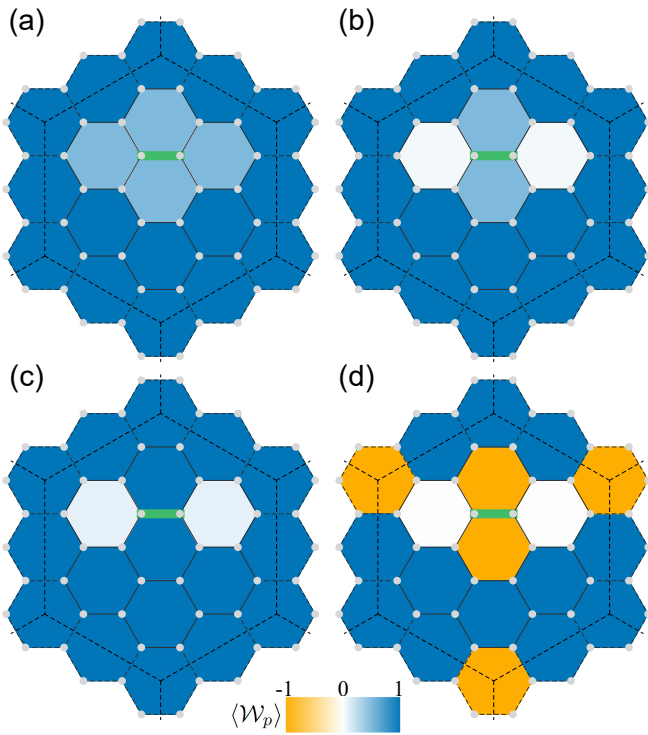} 
\caption{\label{fig:various}Vortex configurations in the ground state when the FM Kitaev interaction $K^z=-1$ on the green $z$ bond is replaced by (a) the FM Heisenberg interaction $J=-1$, (b) the symmetric off-diagonal interaction $\Gamma'=\pm1$, (c) the other symmetric off-diagonal interaction $\Gamma=\pm1$, and (d) the DM interaction $D=\pm1$. The color in each hexagonal plaquette represents the value of the localized vortex $\langle \mathcal{W}_{p} \rangle$. The calculations are done for the 24-site cluster in the dashed hexagon under the periodic boundary condition.}
\end{figure} 

Figure~\ref{fig:various} shows the results for a 24-site cluster with the replacement of the Kitaev interaction $K$ on a $z$ bond (green one in the figure) by (a) $J$, (b) $\Gamma'$, (c) $\Gamma$, and (d) $D$. The color in each hexagonal plaquette indicates the expectation value of $\mathcal{W}_p$ in Eq.~(\ref{eq:Wp}) in the ground state, $\langle \mathcal{W}_p \rangle$. 
Note that in all cases the ground states are doubly degenerate within the precision of $O(\varepsilon)$ and one of them is shown in the figures.

When we introduce the Heisenberg interaction $J$, $\mathcal{W}_p$ for the four plaquettes encompassing the modulated bond are no longer $\mathbb{Z}_2$-conserved quantities, while the others remain conserved; for instance, $J=-1$ gives $\langle \mathcal{W}_p \rangle \simeq 0.734$ for the four plaquette, as shown in Fig.~\ref{fig:various}(a). The value of $\langle \mathcal{W}_p \rangle$ changes as $J$, while it is common to the four plaquette. Meanwhile, when we introduce $\Gamma'$, the change of $\langle \mathcal{W}_p \rangle$ appears on the four plaquettes similarly to the case of $J$, but the value of $\langle \mathcal{W}_p \rangle$ on the two plaquettes including the modulated bond is different from that on the rest two on the side, as shown in Fig.~\ref{fig:various}(b). For $\Gamma$, however, the two plaquettes including the modulated bond remain vortex-free, namely, $\langle \mathcal{W}_{p} \rangle = +1$, as shown in Fig.~\ref{fig:various}(c). This is because the two $\mathcal{W}_p$ commute with the modulated Hamiltonian. 

The most interesting change occurs when we introduce $D$, as shown in Fig.~\ref{fig:various}(d). In this case, similar to $\Gamma$, the two plaquettes including the modulated bond remain $\mathbb{Z}_2$-conserved, but their $\langle \mathcal{W}_{p} \rangle$ are flipped to $-1$, irrespective of the sign of $D$. Thus, the replacement of the Kitaev interaction by the DM interaction induces the vortices in the two plaquettes including the modulated bond. We note that a vortex appears also on a plaquette on the boundary of the cluster, probably due to the finite-size effect. 

The vortex creation by the introduction of $D$ can be regarded as the Kitaev interaction $K$ with the opposite sign to the original one in the background. This is deduced by rewriting the DM interaction as 
\begin{align} 
D(S_i^xS_{i'}^y-S_i^yS_{i'}^x)&=  D\left(-4S_i^yS_i^zS_{i'}^zS_{i'}^x - S_i^yS_{i'}^x \right) \nonumber \\ 
&= 4D S_i^yS_{i'}^x \left( S_i^zS_{i'}^z - \frac14 \right). 
\end{align} 
Our numerical calculations conclude that $D \langle S_i^yS_{i'}^x \rangle>0$ in all cases, which can be regarded as an effective AFM Kitaev interaction on the modulated bond. As the local sign flip of the Kitaev interaction leads to the sign flip of $\langle \mathcal{W}_{p} \rangle$ in the Kitaev model, the vortex creation by $D$ can be understood by this mechanism.  

\begin{figure}
\includegraphics[width=1.0\columnwidth]{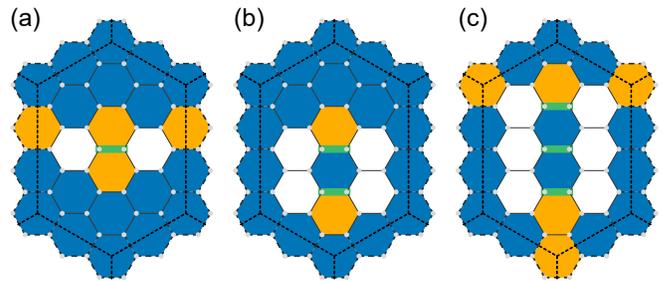} 
\caption{\label{fig:DM32}Vortex configurations in the ground state when the FM Kitaev interactions on the green bonds are replaced by the DM interaction $D=\pm1$. The results are obtained for the 32-site cluster in the dashed hexagon under the periodic boundary condition. In (a), (b), and (c), the one, two, and three bonds are modulated, respectively. The color scale for $\langle \mathcal{W}_{p} \rangle$ is common to Fig.~\ref{fig:various}.}
\end{figure} 

To further examine the effect of the DM interaction, we show the results by similar calculations for a 32-site cluster shown in Fig.~\ref{fig:DM32}. In this case also, the ground states are doubly degenerate. When we introduce $D$ on one bond as shown in Fig.~\ref{fig:DM32}(a), the vortices $\langle \mathcal{W}_{p} \rangle =-1$ are introduced on the two plaquettes including the modulated bond, similar to the 24-site result in Fig.~\ref{fig:various}(d). On the other hand, when we introduce $D$ on two and three $z$ bonds facing with each other as shown in Figs.~\ref{fig:DM32}(b) and \ref{fig:DM32}(c), respectively, the vortices are induced in an interesting manner: $\langle \mathcal{W}_{p} \rangle$ becomes $-1$ on the two plaqeuttes on the ``edges" of the sequence of the modulated bonds, while $\langle \mathcal{W}_{p} \rangle$ remains $+1$ on those sandwiched by the modulated bonds. In other words, the plaquettes including two modulated bonds facing each other remain vortex-free ($\langle \mathcal{W}_{p} \rangle = 1$), while those including only one have vortices ($\langle \mathcal{W}_{p} \rangle = -1$). This behavior is also understood from the effective sign flip of $K$ by $D$ discussed above. The results imply that we can create and move the vortices by successively introducing the DM interactions, as will be demonstrated in Sec.~\ref{sec:ccvortex}. 

Thus far, we discussed the bond modulation starting from the FM Kitaev model with $K<0$. The results for the AFM case with $K>0$ are obtained by using the duality transformation~\cite{CH2010}, where $D$ ($\Gamma$) in the FM case is equivalent to that by $\Gamma$ ($D$) in the AFM case: the vortices are most efficiently induced by $\Gamma$ in the AFM case. 

\subsection{Case study of $\alpha$-RuCl$_3$: ligand displacement} 
\label{subsec:caseRuCl3}

\begin{figure}[ht!]
\includegraphics[width=1.0\columnwidth]{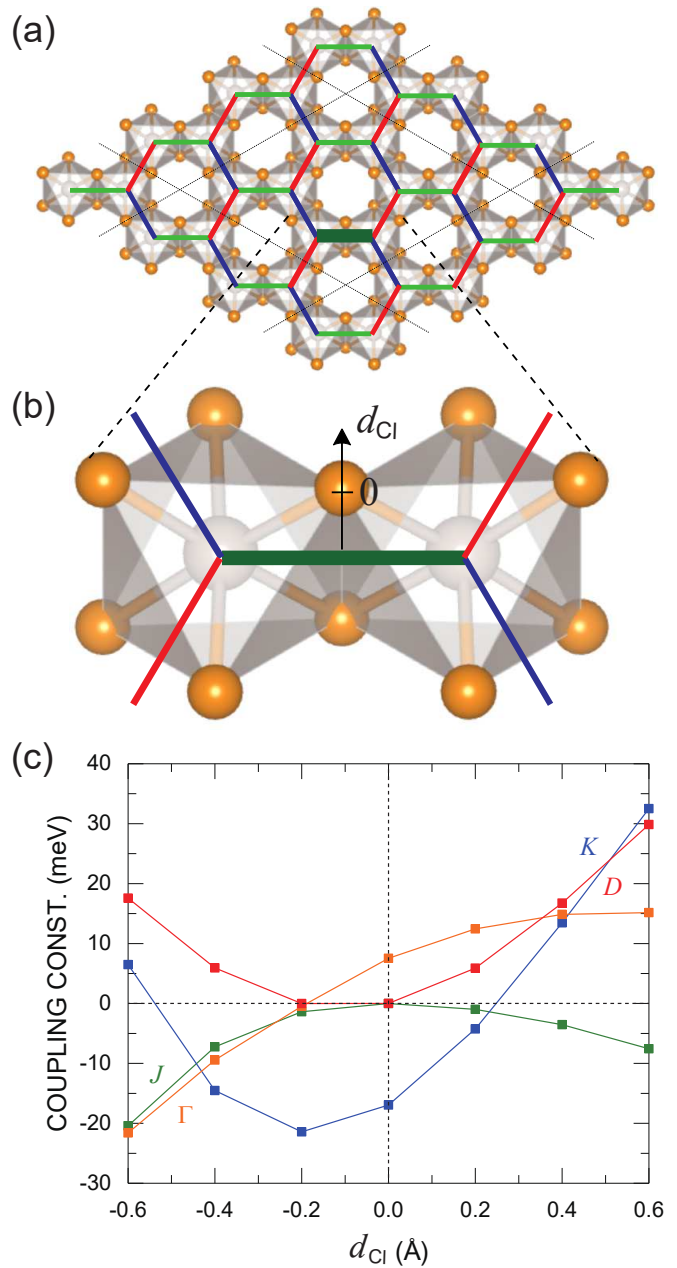} 
\caption{\label{fig:ab-initio}(a) Schematic picture of a monolayer of the Kitaev candidate $\alpha$-RuCl$_3$. The thin dotted lines represent the boundaries for the periodic cells with Ru$_8$Cl$_{24}$. The thick green $z$ bond is modulated by the displacement of the Cl ion with the distance $d_{\rm Cl}$ along the perpendicular direction to the Ru-Ru bond on the plane including the Cl ion and the two Ru cations connected by the Cl, as shown in (b). The atomic positions of the two Ru cations and the Cl ion on the other side of the bond are optimized in the {\it ab initio} calculations. (c) $d_{\rm Cl}$ dependences of the coupling constants of the Kitaev $K$, Dzyaloshinskii-Moriya $D$, Heisenberg $J$, and symmetric off-diagonal $\Gamma$ interactions on the modulated bond. The values are estimated on the basis of the {\it ab initio} band structure for each distorted lattice.
}
\end{figure}

The DM interaction is activated when the spatial inversion symmetry is broken at the bond center~\cite{DZ1958, MO1960}. In candidate materials for the Kitaev model, this can be realized by a local modulation of the lattice structure, e.g., by a displacement of a ligand connecting the magnetic cations. Here, we demonstrate it theoretically for a candidate material $\alpha$-RuCl$_3$ by using {\it ab initio} calculations. We consider a monolayer of this van der Waals material [see Fig.~\ref{fig:ab-initio}(a)], on which a ligand Cl ion is displaced by hand [see Fig.~\ref{fig:ab-initio}(b)]. We compute how the ligand displacement modulates the exchange interactions between the Ru cations by calculating the electronic structure of the modulated $\alpha$-RuCl$_3$ by the {\it ab initio} calculations, constructing the multiorbital Hubbard model with the maximally-localized Wannier functions~\cite{Wannier1937, Marzari1997}, and performing the perturbation expansion in the limit of strong correlation. Similar procedure has been adopted for other Kitaev candidates without bond modulations~\cite{WI2016, Sugita2020, JA2019, JA2020}. 

The {\it ab initio} calculations are performed by using \texttt{Quantum ESPRESSO}~\cite{GI2017}. We adopt the full-relativistic and non-relativistic projector-augmented-wave-method Perdew-Burke-Ernzerhof type for the Ru ions and the Cl ligands, respectively~\cite{BL1994, Perdew1996, DalCorso2014}. The kinetic energy cutoff is set to $200$~Ry. We perform the structural optimization for a monolayer of $\alpha$-RuCl$_3$, whose initial structure within the layer is excerpted from the experimental data for the bulk sample~\cite{JO2015}. The optimization is done for a periodic cell with Ru$_8$Cl$_{24}$ with vacuum space larger than $10$~\si{\angstrom} between the layers. In the structural optimization, we fix the cell vectors and the site positions except for the two Ru cations and one Cl ion involved in the modulated bond; the displaced Cl ion is fixed at the position with a given displacement $d_{\textrm{Cl}}$ along the perpendicular direction to the Ru-Ru bond on the plane spanned by the two Ru cations and the ligand [see Fig.~\ref{fig:ab-initio}(b)]. Note that this procedure modulates the angles of two Ru-Cl-Ru bonds (via the upper and lower Cl) differently. We obtain the optimized structure by setting the minimum ionic displacement to 0.001~\si{\angstrom} in the Broyden-Fletcher-Goldfarb-Shanno iteration scheme~\cite{BR1970}. We take $10\times10\times1$ and $20\times20\times1$ Monkhosrt-Pack $\mathbf{k}$-grids for self-consistent and non self-consistent field calculations, respectively~\cite{MO1976}. We set the convergence threshold for the self-consistent field calculations to $10^{-10}$~Ry. By using \texttt{WANNIER90}~\cite{MO2014}, we construct the maximally-localized Wannier functions for Ru $4d$ $t_{2g}$ and Cl $3p$ orbitals from the result of the electronic band structures, and estimate the effective transfer integrals between the nearest-neighbor Ru $t_{2g}$ orbitals; the scheme is common to Refs.~\onlinecite{JA2019, JA2020}. Given the transfer integrals, we compute the coupling constants for the modulated $z$ bond in Eq.~(\ref{eq:Heff}) by the perturbation expansion for the multiorbital Hubbard model from the strong coupling limit; the scheme is common to Refs.~\onlinecite{WI2016, Sugita2020}. In the expansion, we set the onsite Coulomb interaction $U$, the Hund's-rule coupling $J_{\rm H}/U$, and the spin-orbit coupling coefficient $\lambda$ to $5$~eV, $0.1$~eV, and $0.1$~eV, respectively; the parameters are taken, in consideration of the theoretical and experimental studies for Ru metal~\cite{PhysRevB.83.121101}, Sr$_2$RuO$_4$~\cite{PhysRevB.86.165105}, and $\alpha$-RuCl$_3$~\cite{PhysRevB.93.075144, SI2016}.

Figure~\ref{fig:ab-initio}(c) shows the coupling constants on the modulated bond as functions of the ligand displacement $d_{\textrm{Cl}}$. For $d_{\rm Cl}=0$, the system has the predominant FM Kitaev coupling $K$ and the subdominant positive symmetric off-diagonal coupling $\Gamma$, consistent with the previous studies~\cite{KI2016, HO2017, Sugita2020}. When we introduce the displacement $d_{\rm Cl}$, the inversion symmetry is broken and the DM coupling $D$ is induced as expected. At the same time, other coupling constants are also modulated by the ligand displacement. While the Kitaev coupling $K$ is FM in the absence of the bond modulation, it is increased by the positive $d_{\textrm{Cl}}$ and turns into AFM for $d_{\textrm{Cl}}\gtrsim 0.2$~\si{\angstrom}. For the negative $d_{\textrm{Cl}}$, $K$ is minimized at $d_{\textrm{Cl}}\simeq -0.2$~\si{\angstrom} and increased for smaller $d_{\textrm{Cl}}$. The Heisenberg coupling $J$ is almost zero in the unmodulated case, but induced to be weakly FM for both $d_{\textrm{Cl}}>0$ and $d_{\textrm{Cl}}<0$. The symmetric off-diagonal coupling $\Gamma$, which is positive at $d_{\textrm{Cl}}=0$, is increased (decreased) by the positive (negative) $d_{\textrm{Cl}}$; it changes the sign for $d_{\textrm{Cl}}\lesssim -0.2$~\si{\angstrom}. The other symmetric off-diagonal coupling $\Gamma'$ takes small values (less than $0.1$~meV) for all the cases (not shown). We note that the ligand displacement may yield other exchange interactions not included in Eq.~(\ref{eq:Heff}), but we find that they are also small in the calculated parameter range. We also note that $d_{\textrm{Cl}} \simeq -0.2$~\si{\angstrom} suppresses the coupling constants other than $K$, namely, it makes the bond come close to the pure Kitaev limit. Similar suppression of the non-Kitaev interactions by widening the angle of the Ru-Cl-Ru bond was also discussed for spatially uniform systems~\cite{Nishimoto2016, WI2016}.

On the basis of the {\it ab initio}-based results, we examine the vortex configuration in the effective spin model with the modulated coupling constants. We assume the ligand displacement with $d_{\textrm{Cl}}=0.4$~\si{\angstrom}, which yields $D$ comparable with $K$ and $\Gamma$. Taking a 24-site cluster similar to that used in Fig.~\ref{fig:various}, we consider the spin Hamiltonian with the coupling constants at $d_{\textrm{Cl}}=0$ on all the bonds except for one of the $z$ bonds where the coupling constants are replaced by those at $d_{\textrm{Cl}}=0.4$~\si{\angstrom}. As it is known that the model with dominant $K$ and subdominant $\Gamma$ tends to stabilize a zigzag-type magnetic order in the ground state~\cite{RA2014, RU2019}, we apply the magnetic field to suppress it by adding the Zeeman coupling as
\begin{equation}
\mathpzc{H}_{\rm Zeeman} = - \sum_{i} \mathbf{h} \cdot \mathbf{S}_i. 
\label{eq:Hfield}
\end{equation}
The previous study for $K \simeq -0.94$, $\Gamma \simeq 0.39$, and $\Gamma' = -0.03$ showed that when $\mathbf{h}$ is applied along the $5$~\si{\degree}-deviated direction from $[111]$ towards $[11\overline{2}]$, the zigzag order is suppressed at $|\mathbf{h}|=h \simeq 0.28$, and the Kitaev-like QSL is realized for $0.28\lesssim h\lesssim 0.66$~\cite{Gordon2019}. Employing the same field direction, we perform the exact diagonalization to obtain the vortex configuration in the ground state for the system with bond modulation. 

\begin{figure}[ht!]
\includegraphics[width=1.0\columnwidth]{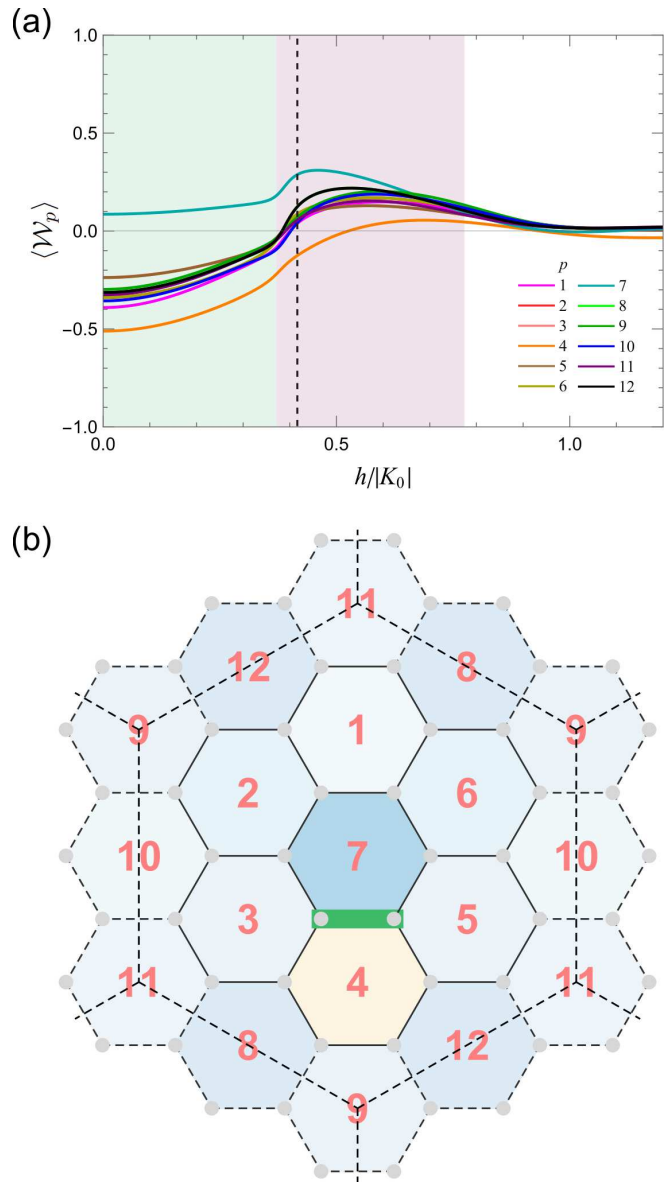} 
\caption{\label{fig:external-h}(a) Magnetic field $h$ dependences of $\langle \mathcal{W}_{p} \rangle$ for twelve hexagonal plaquettes obtained by the exact diagonalization of the 24-site cluster. The numbering of the hexagon, $p$, is shown in (b). The energy unit $K_0$ is taken as the Kitaev coupling constant $K$ at $d_{\rm cl}=0$ in Fig.~\ref{fig:ab-initio}(c). The yellow, orange, and white regions represent the zigzag ordered, the Kitaev-like QSL, and the polarized states, respectively. The dashed line denotes $h/|K_0|=0.42$.  (b) Spatial distribution of $\langle \mathcal{W}_{p} \rangle$ at $h/|K_0|=0.42$ [dashed vertical line in (a)]. The green bond represents the modulated $z$ bond where the coupling constants are set at those for $d_{\rm Cl}=0.4$~\si{\angstrom} in Fig.~\ref{fig:ab-initio}(c); on the rest bonds, the coupling constants are set at those for $d_{\rm Cl}=0$. The color scale for $\langle \mathcal{W}_p \rangle$ is common to Fig.~\ref{fig:various}.}
\end{figure} 

Figure~\ref{fig:external-h}(a) shows $\langle \mathcal{W}_{p} \rangle$ for the twelve hexagons in the 24-site cluster as functions of the magnetic field strength $h$. The numbering of the hexagons are shown in Fig.~\ref{fig:external-h}(b), where the modulated $z$ bond is represented by green. In the presence of the magnetic field, the three ground-state phases are found: the zigzag-ordered state for $0 \leq h/|K_0|\lesssim 0.381$, the Kitaev-like QSL for $0.381 \lesssim h/|K_0|\lesssim 0.774$, and a polarized state for $h/|K_0|\gtrsim 0.774$, where $K_0$ is the Kitaev coupling constant $K$ at $d_{\rm cl}=0$ in Fig.~\ref{fig:ab-initio}(c). The phases are assigned by following Ref.~\onlinecite{Gordon2019}. As shown in Fig.~\ref{fig:external-h}(a), $\langle \mathcal{W}_{p} \rangle$ for different plaquettes change with $h$ in a different manner, especially on the two plaquettes $4$ and $7$ including the modulated bond. The spatial configuration of $\langle \mathcal{W}_{p} \rangle$ at $h/|K_0|=0.42$ is plotted in Fig.~\ref{fig:external-h}(b). We note that it is qualitatively different from the ideal cases with the DM-type modulation in Figs.~\ref{fig:various}(d) and \ref{fig:DM32}(a). This is due to the breaking of $C_2$ rotational symmetry around the center of the target bond by the coexistence of $D$ and $\Gamma$. Interestingly, however, in this region of the Kitaev-like QSL near the lower-field zigzag state, $\langle \mathcal{W}_{p} \rangle$ at the plaquette $4$ takes a negative value, while all the other $\langle \mathcal{W}_{p} \rangle$ are positive. Although the negative value is rather far from $-1$ because of the non-Kitaev contributions mostly from $\Gamma$, our result suggests that the bond modulation by $d_{\textrm{Cl}}=0.4$~\si{\angstrom} can induce a vortex-like object locally in the vortex-free-like background.

\section{Manipulation of vortices} 
\label{sec:ccvortex}

\begin{figure}[ht!]
\includegraphics[width=1.0\columnwidth]{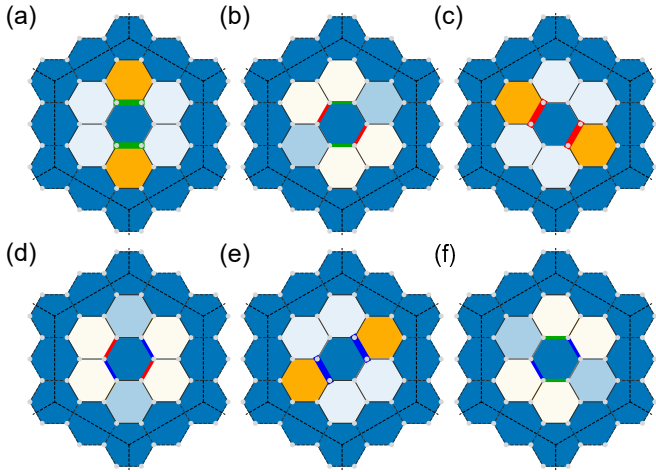} 
\caption{\label{fig:adiabatic_evolution}Braiding of a pair of vortices by adiabatic processes from (a) to (f), and back to (a). The thick green, red, and blue lines represent the $z$, $y$, and $x$ bonds, respectively, on which the DM interactions are imposed. The thickness represents the magnitude of $D$. The color scale for $\langle \mathcal{W}_{p} \rangle$ is common to Fig.~\ref{fig:various}.}
\end{figure} 

\begin{figure}[ht!]
\includegraphics[width=1.0\columnwidth]{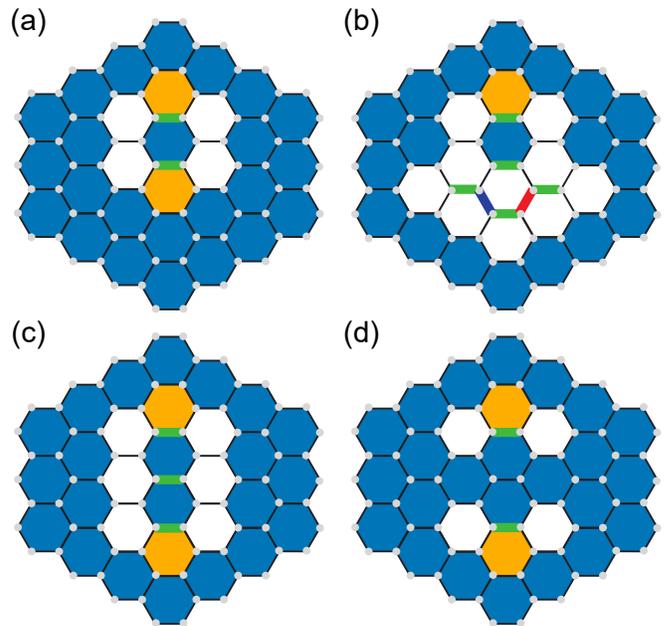} 
\caption{\label{fig:moving-braiding}Adiabatic processes of moving one of a pair of vortices created by the introduction of the DM interactions on the two bonds indicated by green in (a). The notations are common to those in Fig.~\ref{fig:adiabatic_evolution}.}
\end{figure} 

Given the creation of vortices by local bond modulation discussed in Sec.~\ref{sec:lvortex}, in this section, we propose adiabatic processes of controlling the vortices, by successively manipulating multiple bonds. In the following, we consider again the pure FM Kitaev model with bond modulation discussed in Sec.~\ref{subsec:lbond} for simplicity; we believe that similar manipulations are feasible in more realistic cases as discussed in Sec.~\ref{subsec:caseRuCl3}. 

Let us first propose adiabatic processes for a braiding of two vortices. For this purpose, we start with a pair of vortices created by replacing the Kitaev interaction $K$ by the DM interactions $D$ on the two $z$ bonds facing each other, as shown in Fig.~\ref{fig:adiabatic_evolution}(a). Then, we adiabatically weaken $D$ and recover $K$ on the $z$ bonds, and at the same time, gradually replace $K$ by $D$ on the next $y$ bonds, as shown in Fig.~\ref{fig:adiabatic_evolution}(b). By continuing this process until the $z$ bonds recover the original Kitaev interactions while the $y$ bonds are dominated by the DM interactions, we may achieve a $\pi/3$-rotation of the vortex pair, as shown in Fig.~\ref{fig:adiabatic_evolution}(c). Likewise, adiabatically modulating the next $x$ bonds, we may be able to rotate by $\pi/3$ additionally, reaching to Fig.~\ref{fig:adiabatic_evolution}(e) via Fig.~\ref{fig:adiabatic_evolution}(d). Eventually, an additional $z$-bond modulation as in Fig.~\ref{fig:adiabatic_evolution}(f) will end up with a braiding process of the pair of vortices; namely, the vortex configuration is expected to come back to the original one in Fig.~\ref{fig:adiabatic_evolution}(a), but the two vortices are exchanged. 

We confirm that the above braiding process is indeed feasible by using the exact diagonalization for the 24-site cluster of the Hamiltonian with successive adiabatic modulations of the bonds. In the adiabatic modulation, we consider the ``time" $\tau$ dependent Hamitonian given by 
\begin{align}
\mathpzc{H}_{\mu \rightarrow \mu'}(\tau) = (1-\tau) \mathpzc{H}_{\mu} + \tau \mathpzc{H}_{\mu'} + \mathpzc{H}_{\rm Zeeman}, 
\end{align}
where $\mathpzc{H}_{\mu}$ denotes the Hamiltonian in Eq.~(\ref{eq:H_Kitaev}) with the replacement of the FM $K$ by $D$ on two $\mu$ bonds facing each other, leaving the other bonds intact; in $\mathpzc{H}_{\rm Zeeman}$, we set $h=0.005$ along the same field direction as in Sec.~\ref{subsec:caseRuCl3}. In the time evolution, we evolve $\tau$ from $0$ to $1$ with discretization of $\Delta \tau = 10^{-5}$, which always yields the overlap between the eigenstates through the time evolution as $| \langle \tau + \Delta \tau | \tau \rangle | > 0.999$. By applying the time evolution for the sets of bonds in Fig.~\ref{fig:adiabatic_evolution} successively, we find that the expected braiding of two vortices can be achieved. The vortex configurations and the bond thicknesses in Fig.~\ref{fig:adiabatic_evolution} are drawn based on the actual numerical data.

Through the procedures, it is worth noting that the bond modulation by the DM interaction is useful not only to create the vortices with $\langle \mathcal{W}_{p} \rangle=-1$ on the plaquettes adjoining the modulated bond but also to make $\mathcal{W}_{p}$ non-conserved on the plaquettes at both ends of the modulated bonds; see also Fig.~\ref{fig:various}(d) and Fig.~\ref{fig:DM32}. This makes possible to move the vortices to the neighboring non-conserved plaquettes, even for the pure Kitaev model in which the created vortices become conserved.

By extending the above discussion, we can also move vortices apart from each other as follows. Starting from Fig.~\ref{fig:moving-braiding}(a), we can remove one of the vortices by replacing $K$ by $D$ on the five bonds adiabatically, as shown in Fig.~\ref{fig:moving-braiding}(b). By recovering four out of the five as in Fig.~\ref{fig:moving-braiding}(c), we can create another vortex in the lower plaquette, restoring the two plaquettes between the vortices to vortex free. Finally, by removing the DM interaction on the central $z$ bond, one of the vortices in Fig.~\ref{fig:moving-braiding}(a) is moved to the lower plaquette, as shown in Fig.~\ref{fig:moving-braiding}(d). By the reverse procedure, we can also move the vortices toward each other, which would be useful for their fusion.

By combining and modifying the above procedures, we can achieve any creation and move of the vortices. This would serve as a first step to control the fractional excitations toward topological quantum computations.

\section{Summary} 
\label{sec:summary}

In summary, we have theoretically proposed how to engineer vortices in the Kitaev QSL. We showed that the DM ($\Gamma$) interaction can create vortices in the FM (AFM) Kitaev model, as it works as an effective sign-flipper of the Kitaev-type interaction. We demonstrated this mechanism for a Kitaev candidate $\alpha$-RuCl$_3$ monolayer, by introducing a local displacement of the ligand ion. Based on the {\it ab initio} calculations with structural optimization, we found that the displacement induces the DM interaction, which creates a vortex-like feature with negative $\langle \mathcal{W}_{p} \rangle$ in the field-induced Kitaev-like QSL region. In addition, we showed that successive modulations of multiple bonds can create and move the vortices in a designed way, including their braiding and fusion. We expect that its complementary use with other recent proposals with local probe techniques~\cite{PhysRevLett.125.227202, PhysRevLett.126.127201} has a potential for creating, engineering, and detecting vortices experimentally. Our results demonstrate the functionality of the local DM interaction for the control of vortices, which provides a promising way for future topological quantum computation. Our study will stimulate the experimental attempt to achieve anyonic statistics from the topological properties of vortices in the Kitaev QSL, for example, by using the atomic force microscopy.

\begin{acknowledgements}
The crystal structures in Figs.~\ref{fig:ab-initio}(a) and \ref{fig:ab-initio}(b) were visualized by \texttt{VESTA}~\cite{MO2011}. The exact diagonalization with the locally optimal block conjugate gradient method was performed by using ${\mathcal H}\Phi$ package~\cite{KA2017}. Parts of the numerical calculations have been done using the facilities of the Supercomputer Center, the Institute for Solid State Physics, the University of Tokyo. This work was supported by 
Grant-in-Aid for Scientific Research under Grants No.~JP18K03447, No.~JP19H05822, and No.~20H00122  and JST CREST (JP-MJCR18T2).
\end{acknowledgements}


\bibliography{DMflux}

\end{document}